\documentclass{elsart}
\usepackage{graphicx}
\usepackage{rotating}
\usepackage{xspace}

\usepackage{amssymb}

\begin{document}

\begin{flushright}
Fermilab-Pub-06-171-AD\\
\end{flushright}

\begin{frontmatter}




\title{Beam-Based Alignment of the NuMI\\Target Station Components at FNAL}

\author[ut]{R. Zwaska}
\author[bnl]{M. Bishai}
\author[fnal]{S. Childress}
\author[anl]{G. Drake}
\author[unicamp]{C. Escobar}
\author[saopaolo]{P. Gouffon}
\author[fnal]{D.A. Harris}
\author[fnal]{J. Hylen}
\author[ut]{D. Indurthy}
\author[fnal]{G. Koizumi}
\author[ut]{S. Kopp\corauthref{myemail}}
\corauth[myemail]{Corresponding author e-mail {\tt kopp@mail.hep.utexas.edu} } 
\author[fnal]{P. Lucas}
\author[fnal]{A. Marchionni}
\author[fnal]{A. Para}
\author[ut]{\v{Z}. Pavlovi\'{c}}
\author[fnal]{W. Smart}
\author[anl]{R. Talaga}
\author[bnl]{B. Viren}

\address[anl]{Argonne National Laboratory, Argonne, IL}
\address[bnl]{Brookhaven National Laboratory, Upton, Long Island, NY }
\address[fnal]{Fermi National Accelerator Laboratory, Batavia, IL}
\address[saopaolo]{Universidade de Sao Paulo, Instituto de Fisica, Sao Paulo, Brazil}
\address[unicamp]{Universidade Estadual de Campinas, Sao Paulo, Brazil}
\address[ut]{Department of Physics, University of Texas, Austin, TX}

\begin{abstract}
The Neutrinos at the Main Injector (NuMI) facility is a conventional horn-focused neutrino beam which produces muon neutrinos from a beam of mesons directed into a long evacuated decay volume.  The relative alignment of the primary proton beam, target, and focusing horns affects the neutrino energy spectrum delivered to experiments.  This paper describes a check of the alignment of these components using the proton beam.
\end{abstract}

\begin{keyword}
Neutrino detectors, Particle sources and targets,  Beam focusing and bending magnets, Beam monitors
\PACS 
95.55.Vj, 29.25.-t, 41.85.Lc, 41.85.Qg
\end{keyword}
\end{frontmatter}

\section{Introduction}
\label{beamon_intro}

\begin{sidewaysfigure}[p]
  \centering
  \includegraphics[width=8 in]{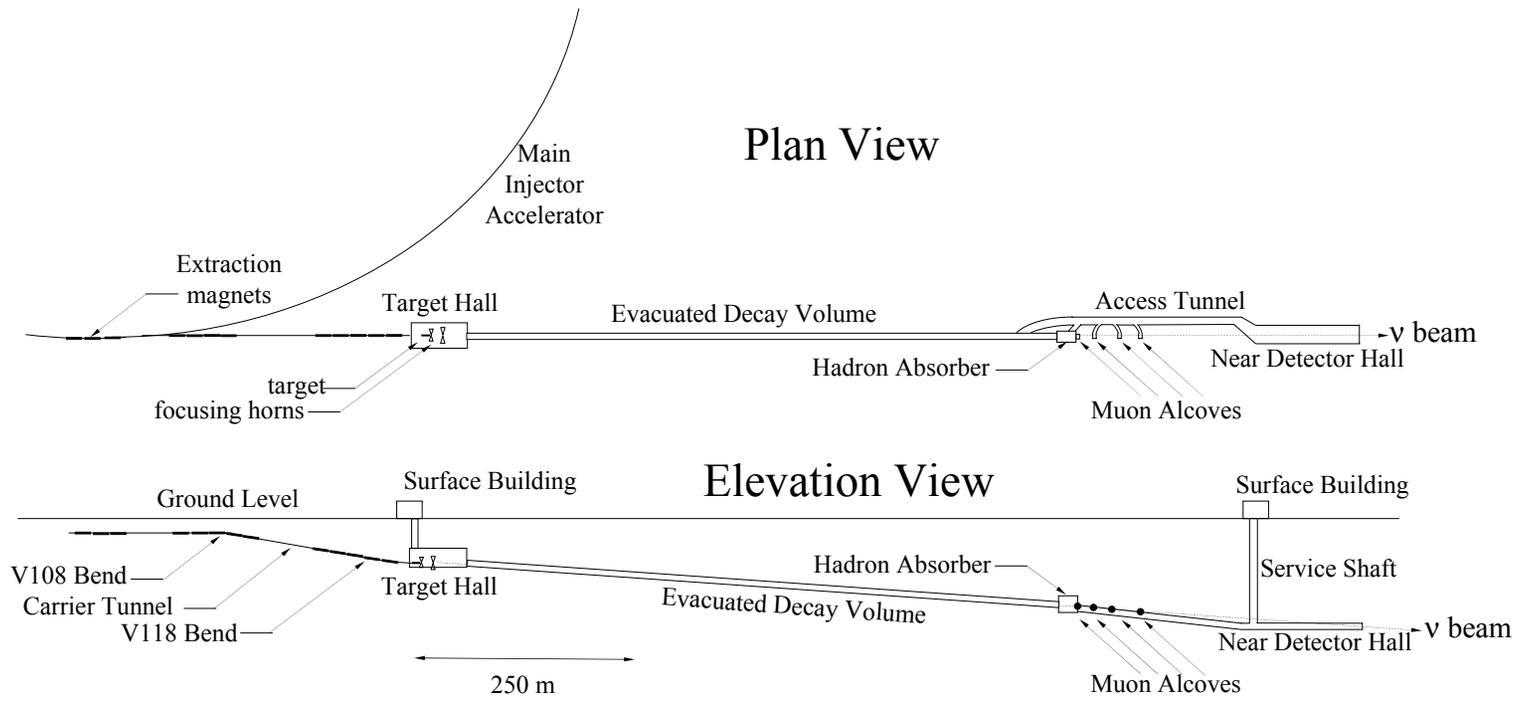}
  \caption{Plan and elevation views of the NuMI beam facility.  A proton beam is directed onto a target, from which the secondary pions and kaons are focused into an evacuated decay volume via magnetic horns.  Ionization chambers at the
end of the beam line measure the uninteracted primary beam, secondary hadron beam and tertiary muon beam.}
  \label{fig:numi}
\end{sidewaysfigure}

\begin{figure}[t]
  \centering
  \includegraphics[width=6 in]{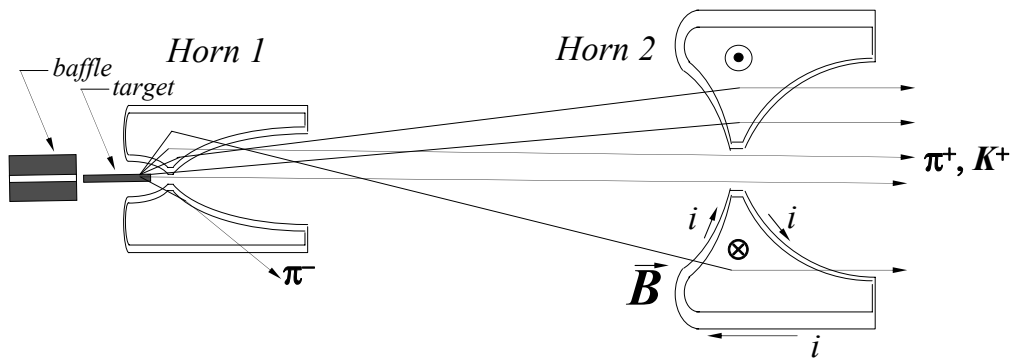}
  \caption{NuMI two-horn beam:  Horns 1 and 2 are separated by $\approx$10~m.  A collimating baffle upstream of the target protects the horns from direct exposure to errant proton beam pulses.  The target and baffle system can be positioned further upstream of the Horns to produce higher energy neutrino beams \cite{flexybeam}.  The vertical scale is $4\times$ that of the horizontal (beam axis) scale.}
  \label{fig:targ-horns}
\end{figure}

The Neutrinos at the Main Injector (NuMI) beam line \cite{numitdr,kopp-numi} at the Fermi National Accelerator Laboratory (FNAL) delivers an intense muon neutrino  beam to the MINOS \cite{minostdr} detectors at FNAL and at the Soudan Laboratory in Minnesota.  Additional experiments \cite{nova,minerva} are being planned.  A scale diagram of the NuMI beam line is shown in Figure~\ref{fig:numi}.
The primary proton beam is fast-extracted from the 120~GeV Main Injector accelerator onto the NuMI pion production target.  The beam line is designed to accept up to $4\times10^{13}$~protons-per-pulse (ppp) with a repetition rate of 0.53~Hz.  After the graphite target, two toroidal magnets called ``horns'' sign-select and focus the secondary mesons from the target, as shown in Figure~\ref{fig:targ-horns}.  The mesons are directed into a 675~m long, 2~m diameter cylindrical volume, evacuated to $\sim0.5$~Torr to reduce pion absorption, where they may decay to muons and neutrinos.  At the end of the decay volume, a beam absorber stops the remaining hadrons, followed by approximately 235~m of unexcavated rock which stops the tertiary muons, leaving only neutrinos.  

\begin{figure}[t]
  \centering
  \includegraphics[width=5. in]{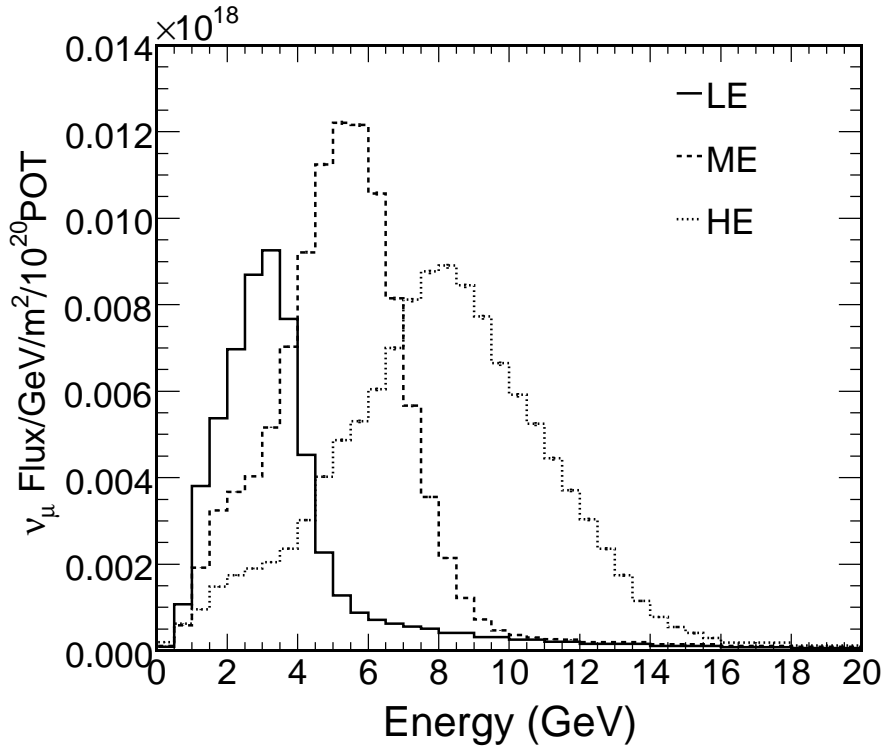}
  \caption{Calculated flux of muon neutrinos in the near detector hall 1040~m from the NuMI target.  Three spectra are shown, corresponding to the low, medium, and high neutrino energy positions of the target \cite{flexybeam}.  In these configurations, the target is located 10, 100, and 250~cm upstream of its fully-inserted position.}
  \label{fig:numi-spectra}
\end{figure}

The target position may be changed remotely so as to produce a variety of wide band beams with peak energies ranging from 3~GeV to 9~GeV \cite{flexybeam}.  The target, shown fully-inserted into the first focusing horn in Figure~\ref{fig:targ-horns}, is mounted on a rail system and can be moved as much as 2.5~m upstream.  Moving the target upstream directs smaller-angle, higher-momentum particles into the magnetic fields of the focusing horns, resulting in a higher-energy neutrino beam, as shown in Figure~\ref{fig:numi-spectra}.\footnote{For maximal efficiency of the ME and HE beams, both the target and the downstream horn are moved with respect to the fixed first horn \cite{numitdr}.  Because of the complexity of moving Horn 2, the MINOS experiment makes use only of the target motion system, which can be accomplished {\it in situ} \cite{flexybeam}.}

\begin{figure}[t]
\centering
\vskip -.8cm
\includegraphics[width=12cm]{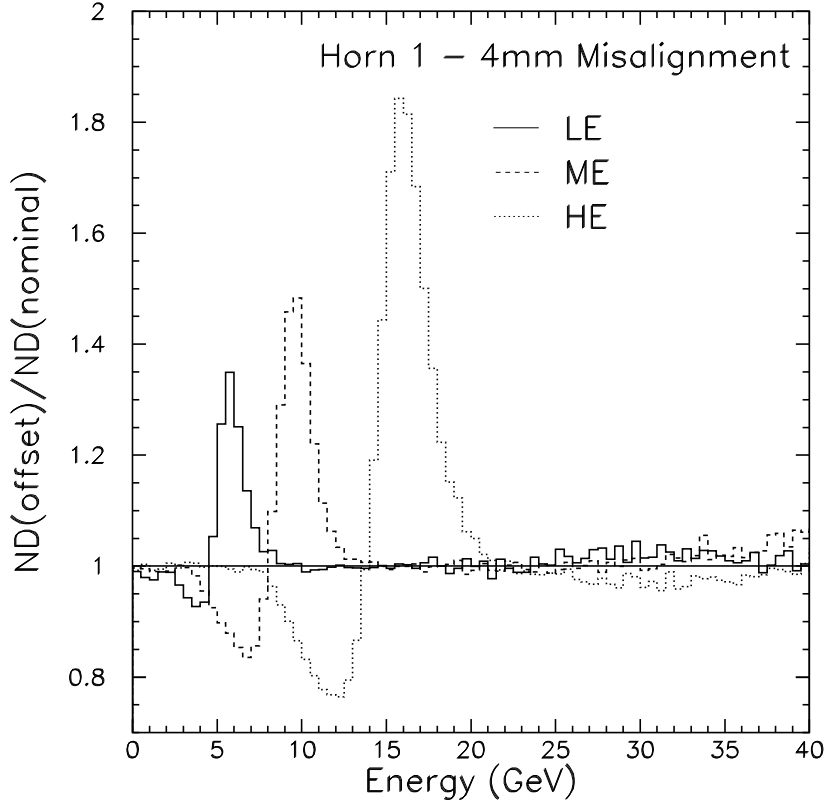}
\vskip -.4cm
\caption{Results of a Monte Carlo simulation showing the effects
of transverse displacements of Horn 1 by 4~mm on the neutrino beam
spectrum at the MINOS near detector.  Plotted is the ratio of
neutrino flux with the offsets compared to the nominal flux.
The misalignment effect is shown for the beam configured with the target in the
LE, ME and HE positions.  }
\label{fig:bmeas_h1off}
\end{figure}

\begin{sidewaysfigure}[p]
  \centering
  \includegraphics[width=8.5 in]{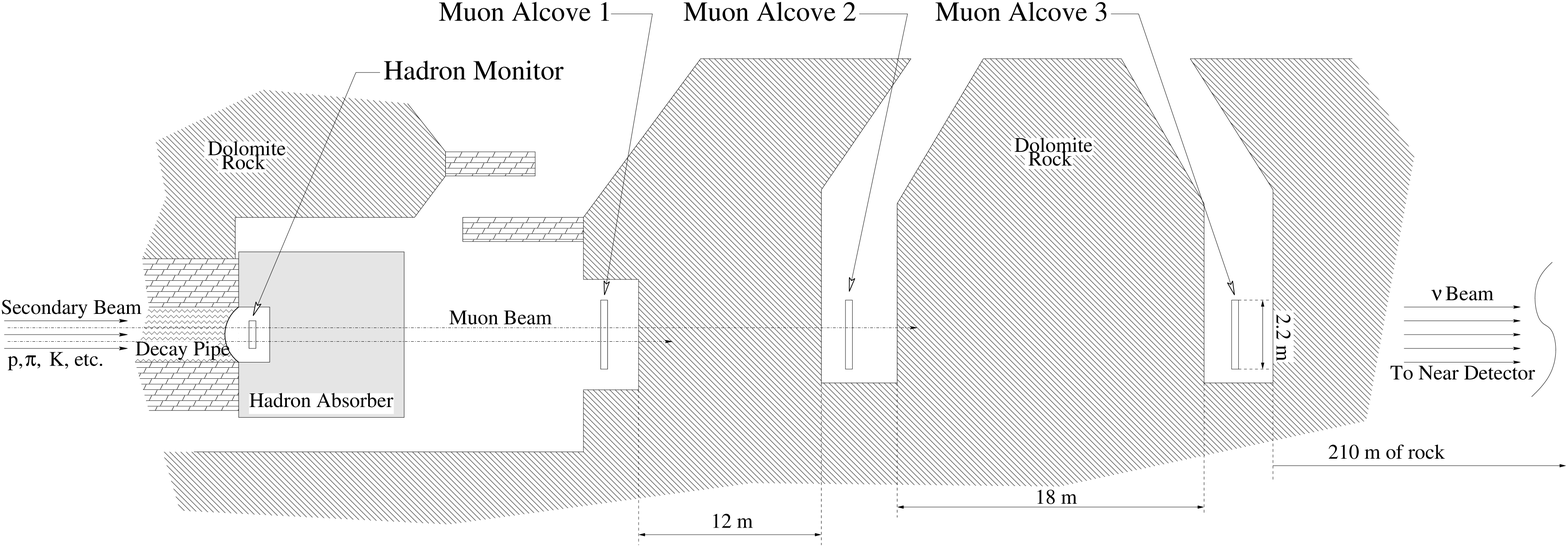}
  \caption{Plan view of the downstream areas of the NuMI beamline.
    The beam, consisting of hadrons, muons, neutrinos, and remnant protons, enters the area through the decay pipe.  The hadron
    beam's spatial distribution is measured at the Hadron Monitor and then
    stopped in the Hadron Absorber.  The higher-energy muons penetrate the
    absorber and some distance of rock; along the way their spatial 
    distributions are measured by the Muon Monitors.  }
  \label{fig:beamon_layout}
\end{sidewaysfigure}

The neutrino energy spectrum, and the ratio of spectra at the the near and far detectors, are sensitive to misalignments of either the target or horns with respect to the beam axis defined by the proton beam, as has been shown in previous experiments \cite{Casagrande}.  The low-energy neutrino beam is particularly sensitive to misalignments because higher energy hadrons will be better focused for almost any misalignment.  For example, an off-center target relative to the proton beam results in fast $\pi/K$ being more able to exit the target without reinteraction in the target material, which ordinarily produces lower-energy mesons. An off-center horn, on the other hand, results in the horn inner conductor being able to intercept and focus smaller-angle mesons off the target, which are the most energetic.  The effect of a transverse offset of Horn 1 on the neutrino energy spectrum in the MINOS near detector ($L=1040$~m from the NuMI target), for example, is shown in Figure~\ref{fig:bmeas_h1off}.

Prior to initial running of this beam line, an optical survey was performed of the primary beam components, target station components, and decay tunnel axis.  The optical alignment of the primary beam and target station components was accurate to $\pm0.5$~mm.  The optical alignment of the cavern housing the beam absorber, 723~m from the target, was accurate to $\pm2.5$~cm.  Beam instrumentation placed in front of the absorber, therefore, can provide a check of the optical alignment of the NuMI beam axis to within $\pm30~\mu$m.  

\begin{figure}[p]
  \centering
  \vskip -1.cm
  \includegraphics[width=13.5cm]{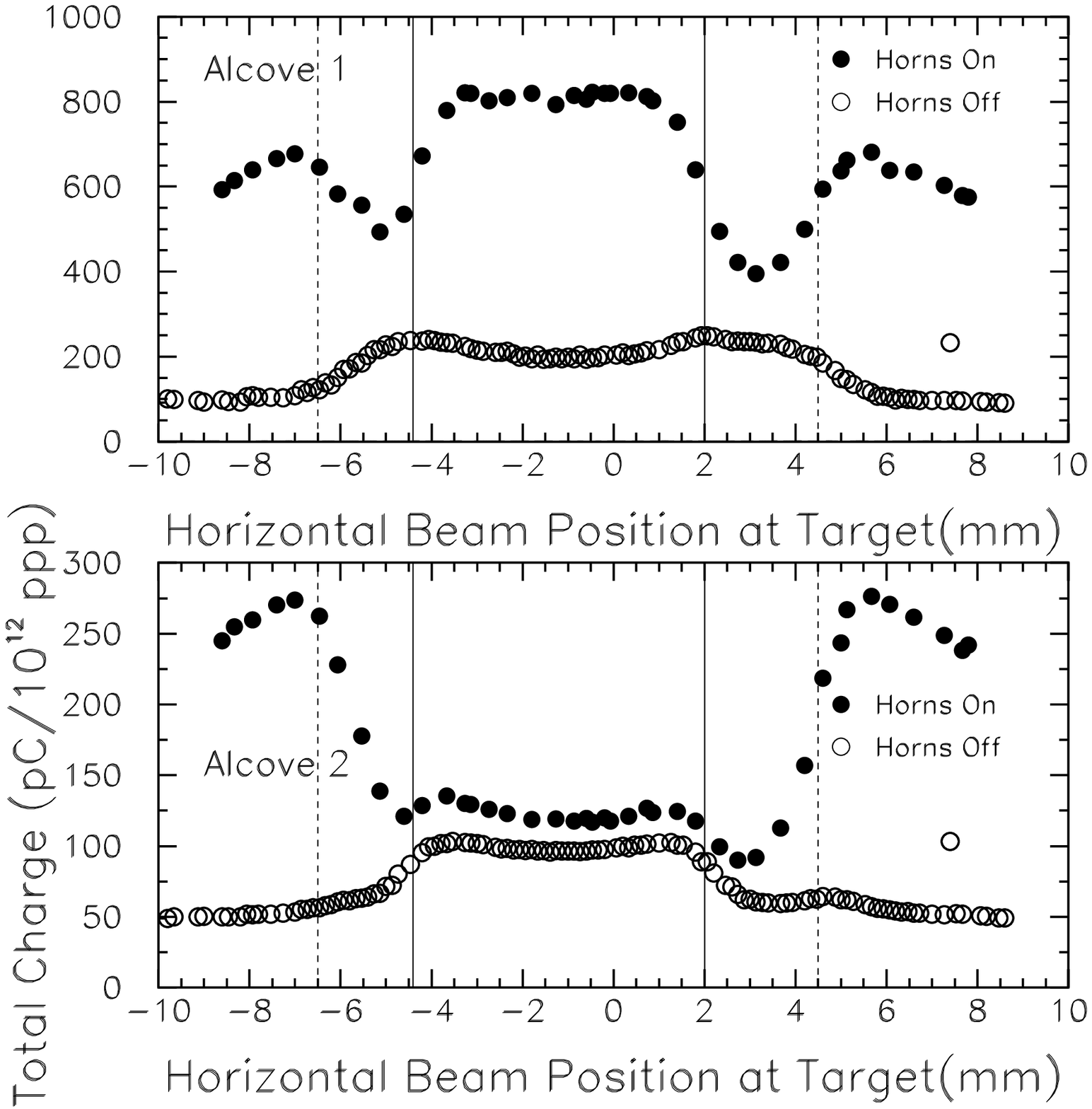}
  \vskip -.7cm
  \includegraphics[width=13.5cm]{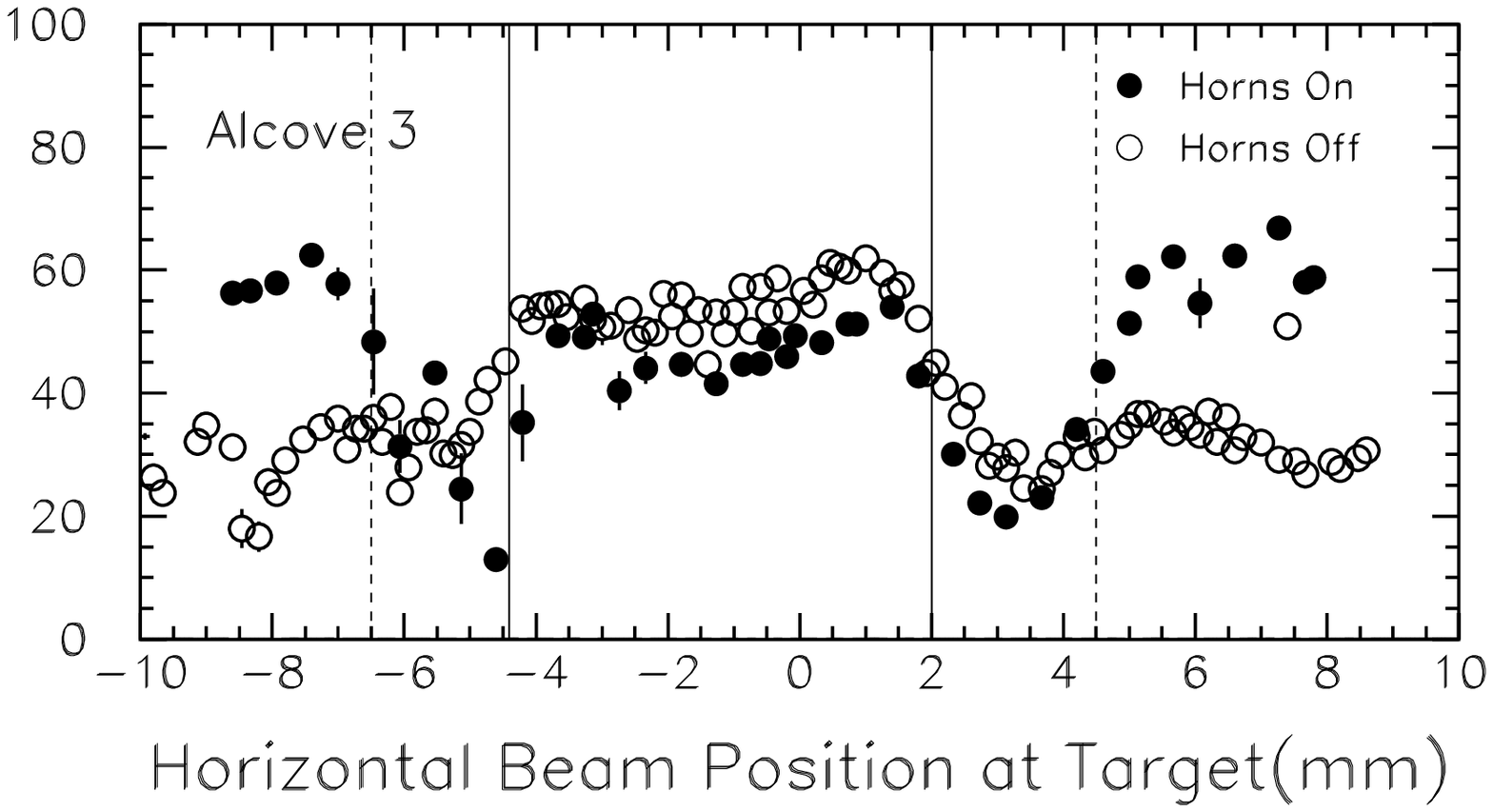}
  \vskip -6.cm
  \caption{Signals from the 3 NuMI Muon Monitor stations during a horizontal scan of the proton beam across the target.
    As determined from the beam scans described later in this paper, the target is centered at $-1.2$~mm, and 
    the solid lines indicate its inferred edges.
    A graphite baffle upstream of the target has an inner aperture of 11~mm diameter, 
    whose edges are indicated by the dashed lines.
    The ``zero position'' of the proton beam was taken from the optical alignment.}
  \label{fig:le-muon-scan}
\end{figure}

In this article we describe the process of beam-based alignment, where the proton beam itself is used to locate the relative positions and angles of these components.  Such a procedure is an independent means of checking the alignment of the beam components done previously by optical survey.  The beam-based alignment avoids the error that would result in the case of our low-energy neutrino beam of aligning the components based on maximal neutrino yield:  as noted above, misalignments tend to produce a higher mean energy neutrino spectrum (by as much as 5\% of the focusing peak energy).  Alignment based on muon yield was performed in some previous experiments ({\it e.g.} \cite{Casagrande,Astier}), but as can be seen from Figure~\ref{fig:le-muon-scan} the largest muon yields in the NuMI LE beam do not arise from protons striking the center of the target.  Thus, a detailed beam-based alignment is essential for low-energy neutrino beams.  A beam-based alignment has been performed previously at Argonne National Laboratory (ANL), by temporarily placing glass plates before and after the target and horns to observe the location of the radiation-blackening \cite{Barish1977}.

This article procedes as follows.  Section~\ref{fiducials} describes the geometry of the NuMI target and horns, especially those geometric features used to locate them with the beam.  Section~\ref{instrumentation} describes the instrumentation used to measure the proton beam position and direction, as well as observe the downstream interactions of the proton beam in the target and horn material.  Section~\ref{target} describes the alignment of the target and collimating baffle.  Section~\ref{horns} describes the alignment of the focusing horns.  Section~\ref{summary} discusses uncertainties on the neutrino flux from these misalignments.

\section{Alignment Fiducials}
\label{fiducials}

\begin{figure}[tbp]
  \centering
  \vskip -.8cm
  \includegraphics[width=13cm]{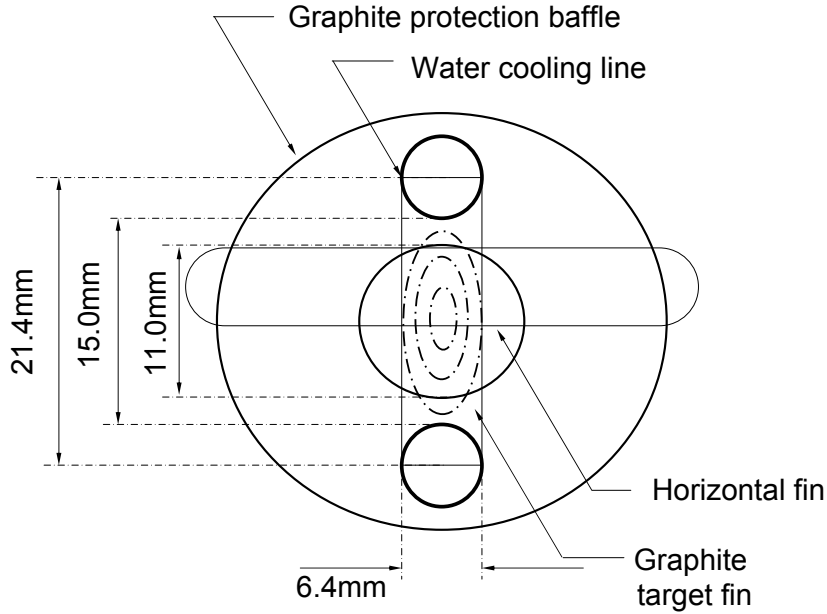}
  \caption{Beam's-eye view of the target-baffle system.  The beam
    first passes through the baffle aperture, then a single horizontal graphite target 
    fin, and then the target itself (47 fins).  The approximate $1, 2, 3~\sigma$ sizes of the proton
    beam during the alignment survey are indicated by the dashed contours.}
  \label{fig:bmeas_tlay}
\end{figure}

\subsection{Target and Baffle System}
\label{targ-fiducials}
A beam's eye view of the target-baffle system \cite{numitdh} is shown in Figure~\ref{fig:bmeas_tlay}.  The target and baffle can be manipulated remotely in all three coordinates as well as pitch and yaw.  The target and baffle are fixed relatively with respect to each other and move as a unit.  

The baffle is a graphite rod with a 11~mm diameter clear aperture. The baffle is 150~cm long, corresponding to 3.1~nuclear interaction lengths and 6.3~radiation lengths.  When the beam is incident on the baffle material (and not the aperture) we expect 4.5\% of the proton beam to survive, and to have acquired a RMS angle of 0.30 mrad due to multiple scattering.  Both effects can be measured using the downstream instrumentation.

Beginning 68~cm downstream of the downstream end of the baffle, the target consists of forty-seven 2~cm long graphite segments, spaced by 0.3 mm.  The segments have a width of 6.4~mm and a height of 15~mm.  The graphite segments comprise, in total, 2.0~nuclear interaction lengths and 4.0~radiation lengths.  When traversing the entire target length, 13.5\% of the proton beam will emerge with a RMS angle of 0.24~mrad due to multiple scattering.  As can be seen in Figure~\ref{fig:bmeas_tlay}, the beam can be steered to the immediate right or left of the target such that it still passes through the baffle aperture. This allows determination of the horizontal target position.

To determine the vertical target position there is a 48$^{\rm th}$ target fin, oriented horizontally, 15 cm upstream of the main target.  This horizontal fin is made of the same graphite material as the main target.  It is centered ($2.26\pm0.25)$~mm above the vertical center of the main target, as determined by post-assembly optical survey.  The horizontal fin contributes 0.08~radiation lengths, measureable as an increase of the proton beam RMS scattering angle.

\subsection{Horns and Crosshairs}
\label{bmeas_bba_chair}

\begin{figure}[t]
  \centering
  \vskip 0.cm
   \includegraphics[width=13.8cm,angle=0]{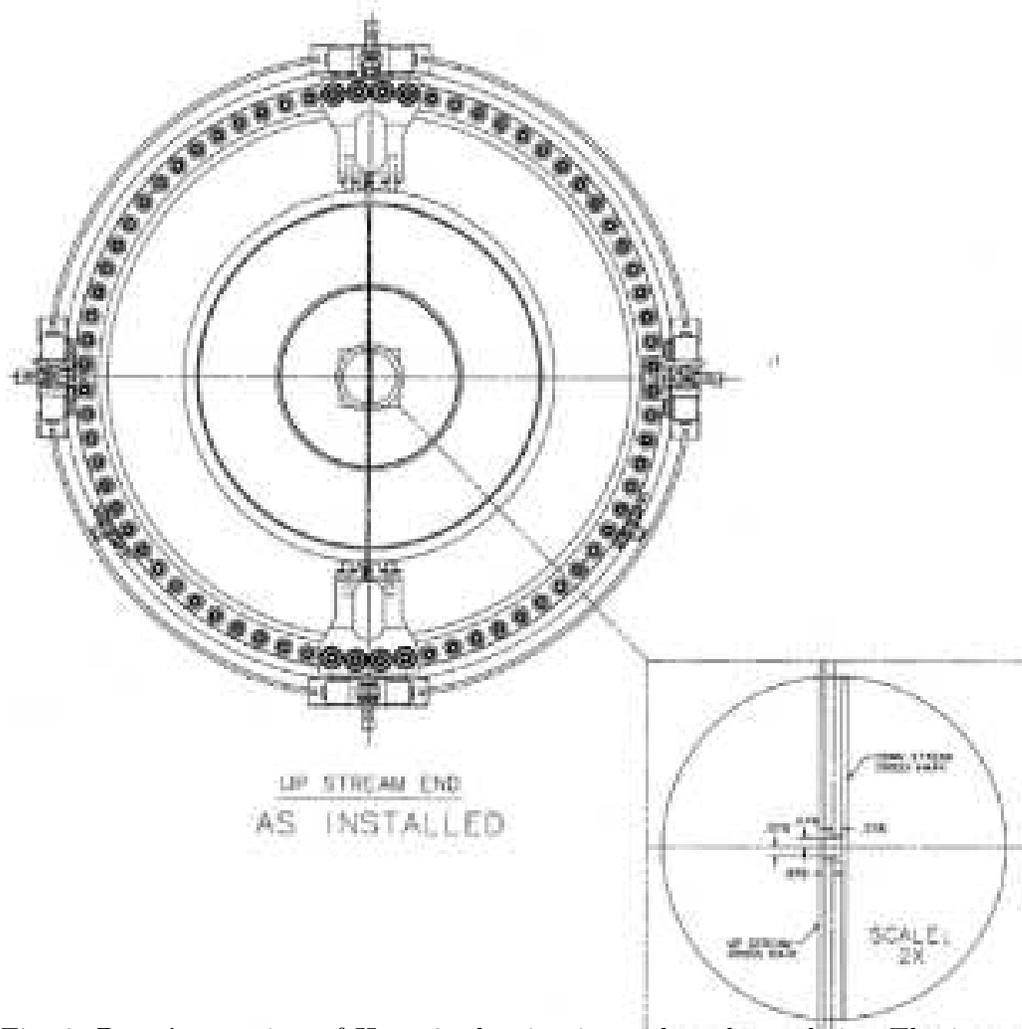}
 \vskip -.9cm
  \caption{Beam's eye view of Horn 2, showing its neck and crosshairs.  The inset at lower-right shows a close-up view of the region near the beam axis, showing the cross-hair mounted to the
    upstream end of Horn~2 at beam left and the downstream cross-hair at beam right.  Each cross hair has a small horizontal segment (a ``nub'') which intercepts the proton beam if the beam is
    steered vertically high or low.  Horn~1, whose neck is just 9~mm in radius, has only a downstream cross-hair at beam left. All dimensions are in inches.}
  \label{fig:bmeas_hlay}
\end{figure}

\begin{figure}[!] 
  \centering
  \includegraphics[width=12cm]{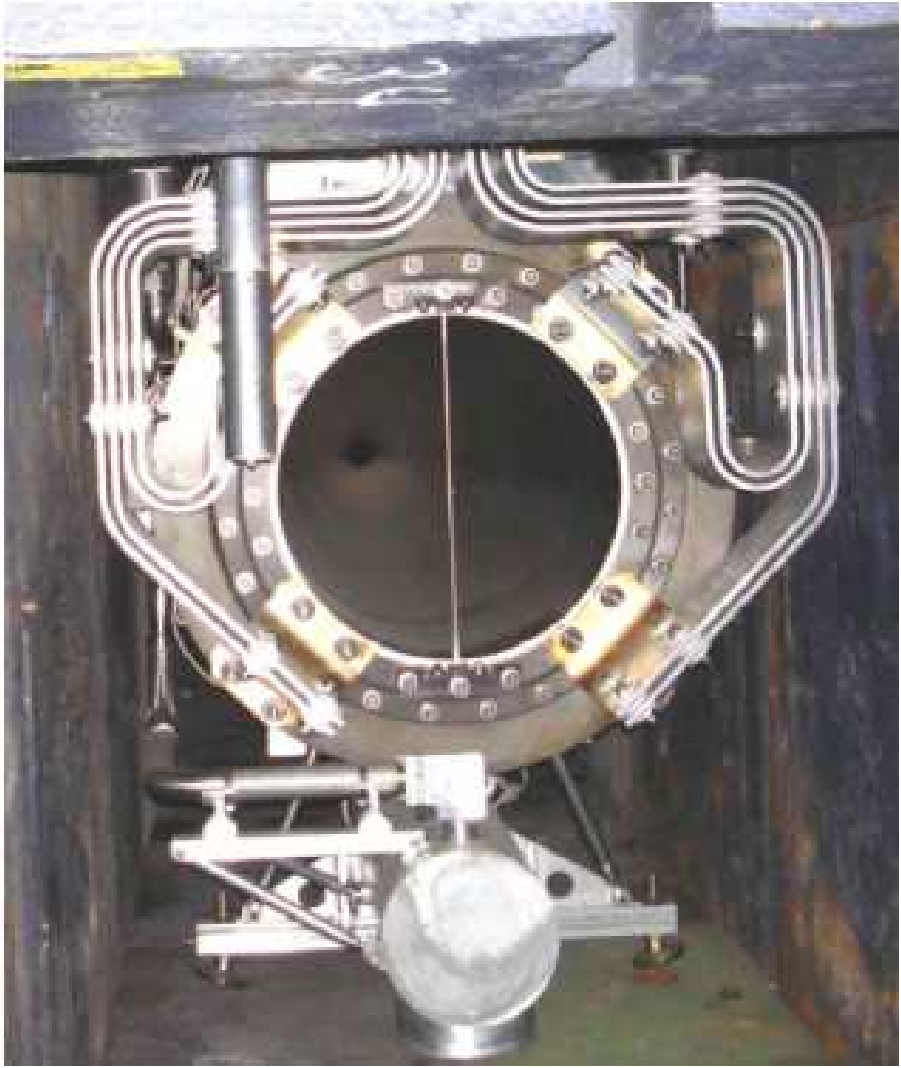}
  \caption{Picture of Horn 2, as installed, from the downstream end.  The 
    crosshair is the vertical bar across the horn aperture. The horizontal 
    ``nub'' on the crosshair is barely visible at the crosshair's center.  The loss
    monitor ion chamber is the cylinder extending from above on the
    left (beam right).}
  \label{fig:bmeas_hornpic}
\end{figure}

To allow beam-based alignment of the horns, a system of ``crosshairs'' was designed \cite{numitdh,Para2001}.  Using these crosshairs and the Horn 1 inner aperture (the horn ``neck'') the upstream and downstream locations of each horn can be independently measured, giving a measurement of both position and angle of the horn.  Figures~\ref{fig:bmeas_hlay} and \ref{fig:bmeas_hornpic} show the crosshairs on the horns.  Such measurements were performed prior to the installation of the target-baffle system, which prevents the proton beam from being able to strike the crosshairs.

The crosshairs are aluminum bars attached to the downstream face of Horn~1, and both the upstream and downstream faces of Horn~2.  Each crosshair consists of a main spar oriented vertically and 
which is 1~mm wide in the horizontal direction.  Each crosshair has an additional horizontal ``nub'' that is 1~mm in vertical height and extends horizontally 3.5~mm toward the center of the beamline (see Figure~\ref{fig:bmeas_hlay}).  In the beam's-eye view the Horn 1 downstream and Horn 2 upstream crosshairs overlap.  By locating the upstream and downstream crosshairs in both the horizontal and vertical directions, the horns' locations and angles with respect to the beam axis can be determined.

Under normal beam operations, the target is in the beam line, and the proton beam does not intercept any of the crosshairs or their extension nubs.  However, with the target removed, the proton beam can be translated horizontally or vertically, and made to cross these features, generating a spray of charged particles caused by nuclear interactions of the proton beam in the crosshair material.\footnote{The crosshairs range from 6~mm to 18~mm in length along the beam direction, corresponding to 1.5\%-4.6\% of a nuclear interaction length.}  Such particle showers can be detected downstream of each horn.  Additionally, the crosshairs induce a 0.03$-$0.05~mrad broadening of the proton beam due to multiple Coulomb scattering in the crosshair material.  

Because the NuMI target penetrates the upstream parabolic end of Horn 1, no upstream crosshair can be mounted on Horn~1.  To provide a measurement of its upstream end, proton beam is scanned until the inner aperture (the ``neck'') of Horn 1 is struck by the proton beam.  The Horn~1 neck is at 9~mm radius.  It is $\sim$4 cm long, providing 10\% of a nuclear interaction length and 45\% of a radiation length, so its effects on the proton beam are easily measurable either from attenuation or increased angular spread.

\section{Instrumentation}
\label{instrumentation}

A pair of ionization chambers, one immediately downstream of each horn, is used to measure the spray of particles created when the proton beam strikes the crosshairs.  One ionization chamber is shown in Figure~\ref{fig:bmeas_hornpic}.  The ionization chambers (also referred to as a ``loss monitors'') are normally retracted out of the beam, but can be lowered into position for the beam-based alignment procedure.  The positions of the loss monitors, known to a couple centimeters, are not particularly important for the alignment procedure;  they provide only a measurement of the relative increase in charged particle fluence.

At the downstream end of the NuMI decay pipe, 723 m from the target station, the Hadron Monitor measures the proton beam profile and intensity (see Figure~\ref{fig:beamon_layout}).\footnote{During normal operation of the neutrino beam, the Hadron Monitor is exposed to protons which do not interact in the target, as well as mesons and other particles reaching the end of the decay tunnel.  However, the proton fluence dominates.  During many of the alignment scans in this paper, the target is removed from the beamline, so only protons reach the Hadron Monitor.}  The Hadron Monitor is an array of 49 ionization chambers at the end of the decay pipe, and is described further in Ref.~\cite{beamon-paper}.  The Hadron Monitor's measurement of the proton beam relative intensity is stable to better than 1\% over the 3$-$4~hour periods of the alignment scans.  The 49 chambers of the Hadron Monitor provide 49\% coverage \footnote{One chamber with erratic signal was excluded, so the coverage is only 48\%.} of an active square of side 76.2 cm for an angular coverage of 1.06 mrad in each plane ($\pm$ 0.53 mrad).

The target fins are electrically isolated, so can be read out as a ``Budal'' monitor \cite{budal}.  In such a technique, the charge ejected from the target when struck by the proton beam can be read by charge-sensitive electronics, the charge ejected being proportional to the fraction of the proton beam striking the target.  Such a technique can provide a measurement of the target's position relative to the proton beam position, as was demonstrated in a beam test of a prototype target \cite{numitarget}.

The proton beam position at the target hall is extrapolated based on instrumentation upstream of the target station.  Two instrumentation stations, located just after the last focusing quadupole (referred to as the ``121'' instrumentation station), and a second at the end of the primary beam transport line (referred to as the ``TGT'' instrumentation station) are used to extrapolate the proton beam position.  There are no magnets in between the two instrumentation stations, permitting an extrapolation to the target with a simple linear model:
\begin{equation}
  x(z) = x_{\rm TGT} + (x_{\rm TGT} - x_{\rm 121}) \times \left ( 
  \frac{z - z_{\rm TGT}} {z_{\rm TGT} - z_{\rm 121}} \right )
  \label{bmeas_bba_ext}
\end{equation}
$x_{\rm 121}$ and $x_{\rm TGT}$ are the measured proton beam positions at the 121 and TGT instrumentation stations (in either the horizontal or vertical views).  The locations $z_{\rm 121}$ and $z_{\rm TGT}$ are the positions of the instrumentation stations along the beamline, known from survey; $x(z)$ is the extrapolated position of the proton beam at the longitudinal location $z$.   For reference, the two instrumentation stations are approximately 12~m apart.  The target is approximately 10~m downstream of the TGT station, while Horn~2 is approximately 33~m from the TGT instrumentation station.  

Each of the above instrumentation stations consists of a segmented foil secondary emission monitor (SEM) which provides horizontal and vertical beam position and profile measurements \cite{biw04}.  The precision of the SEM position measurements is approximately $25~\mu$m for the low proton beam intensities used during the alignment scans \cite{sem-beamtest}.  The extrapolated resolutions from SEM measurements are estimated to be $\sim$ 32~$\mu$m for the baffle, target, and Horn 1; and 54 $\mu$m at Horn 2.  Each instrumentation station also has a horizontal and a vertical capacitative Beam Position Monitor (BPM) which are used during normal NuMI beam operations \cite{prieto}.  Because of the low proton beam intensities during the alignment scans, the resolution of the extrapolated beam position using the BPMs were 260~$\mu$m and 450~$\mu$m at the two locations.  The BPMs, furthermore, required knowledge of the electrical centers, which introduces an overall ``zero position'' error from the BPMs in addition to the optical alignment of their mechanical locations in the beam line.  Thus, the SEM measurements were used for the beam-based alignment measurements because of their better resolution at low intensity.\footnote{During normal beam operations, the BPMs are used to steer the proton beam.  Therefore, the target and horn positions are also given at the end of this article in BPM coordinates.}

The proton beam intensity is determined by a toroid located at the TGT instrumentation station.  During normal NuMI beam operations, the beam intensity is $(2-3)\times10^{13}$~protons-per-pulse (ppp), although during the beam-based alignment meausurements the beam intensity was reduced to $(4-8)\times10^{11}$ppp.  The toroid precision is $\approx10^{10}$~ppp.  

Corrector dipoles (19 in total) upstream of the last quadrupole can provide lateral displacements of the proton beam up to $\pm12$~mm.  The beam can be displaced without introduction of an angle, {\it i.e.} the displacements retain the beam parallel to the nominal beam axis.  Such lateral, parallel beam displacements are accomplished by tuning of three upstream corrector dipole magnets in pre-determined proportionally-adjusted currents.  The two instrumentation stations, 121 and TGT, monitor to what extent the proton beam develops an angle during scans across the target and horns.

Prior to installation of the target, the proton beam was transported through the primary beam line and permitted to drift through the target hall to the hadron monitor and beam absorber.  The primary beam SEMs could be used to predict the proton beam position 723~m downstream at the location of the Hadron Monitor.  The agreement between this prediction and the actual proton beam position measured by the Hadron Monitor was better than 2~cm in both the horizontal and vertical directions.  This measurement confirms the absolute proton beam axis established by the optical alignment to within $30~\mu$rad.

\section{Alignment of Target and Baffle}
\label{target}

The target and baffle horizontal and vertical positions can be established by horizontal and vertical scans of the target.  The angles of the target and baffle cannot be well measured.  

The target alignment was performed on March 3, 2005, and April 25, 2005, the second alignment necessitated by the target's removal and re-insertion in the beam line.  The alignment scans performed before and after this occaision indicate that the target's position shifted by $\approx0.5$~mm was noted between these two occasions.  During the April 25 scans, the proton beam profile at the target was $\sigma_x \times \sigma_y = 0.9 \times 0.9$~mm$^2$, while for the March~3 scans the spot size was skewed due to an incorrect quadrupole setting:  $\sigma_x \times \sigma_y =0.7\times 1.4$~mm.

\subsection{Vertical Target-Baffle Measurements}
\label{bmeas_bba_tp_v}

\begin{figure}[t]
\centering
\vskip -.8 cm
\includegraphics[width=13.5cm]{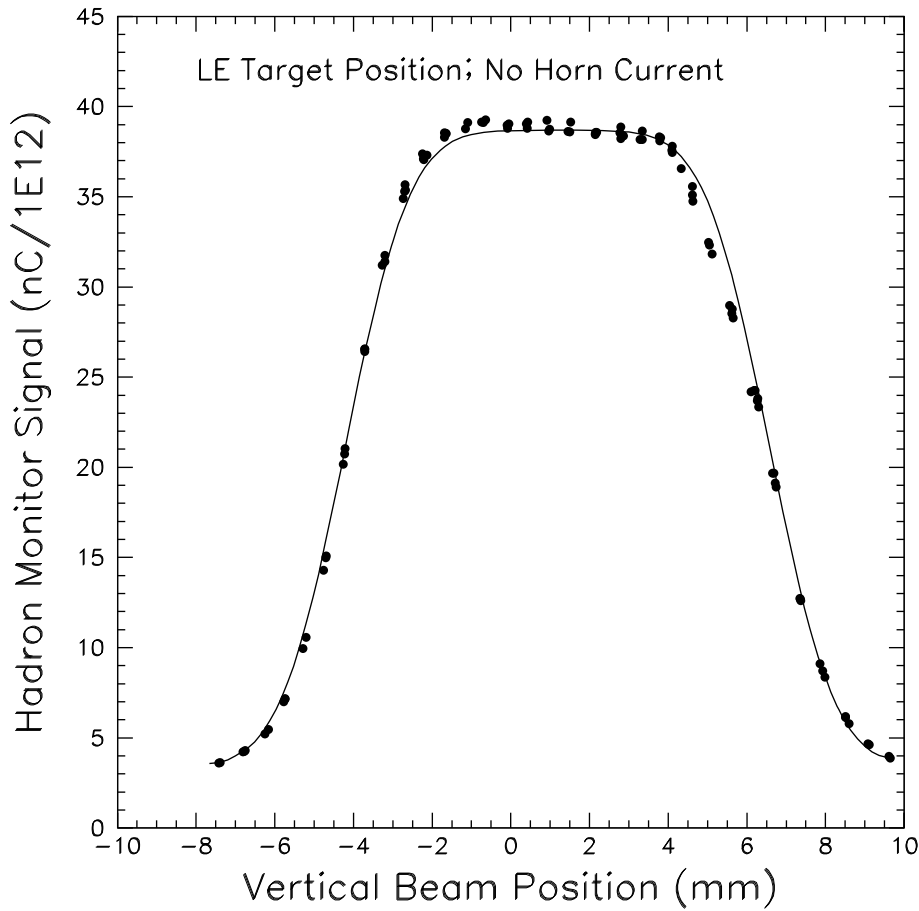}
\caption{Vertical scans of the proton beam across the target and baffle
performed on March~5, 2005.
Plotted is the total amount of charge collected in the Hadron Monitor ion chamber array, 
normalized by proton beam intensity, and plotted as a function of
proton beam position at the target.  }
\label{fig:bmeas_vtscans} 
\end{figure}

\begin{figure}[t] 
  \centering
  \vskip -.8 cm
  \includegraphics[width=13.5cm]{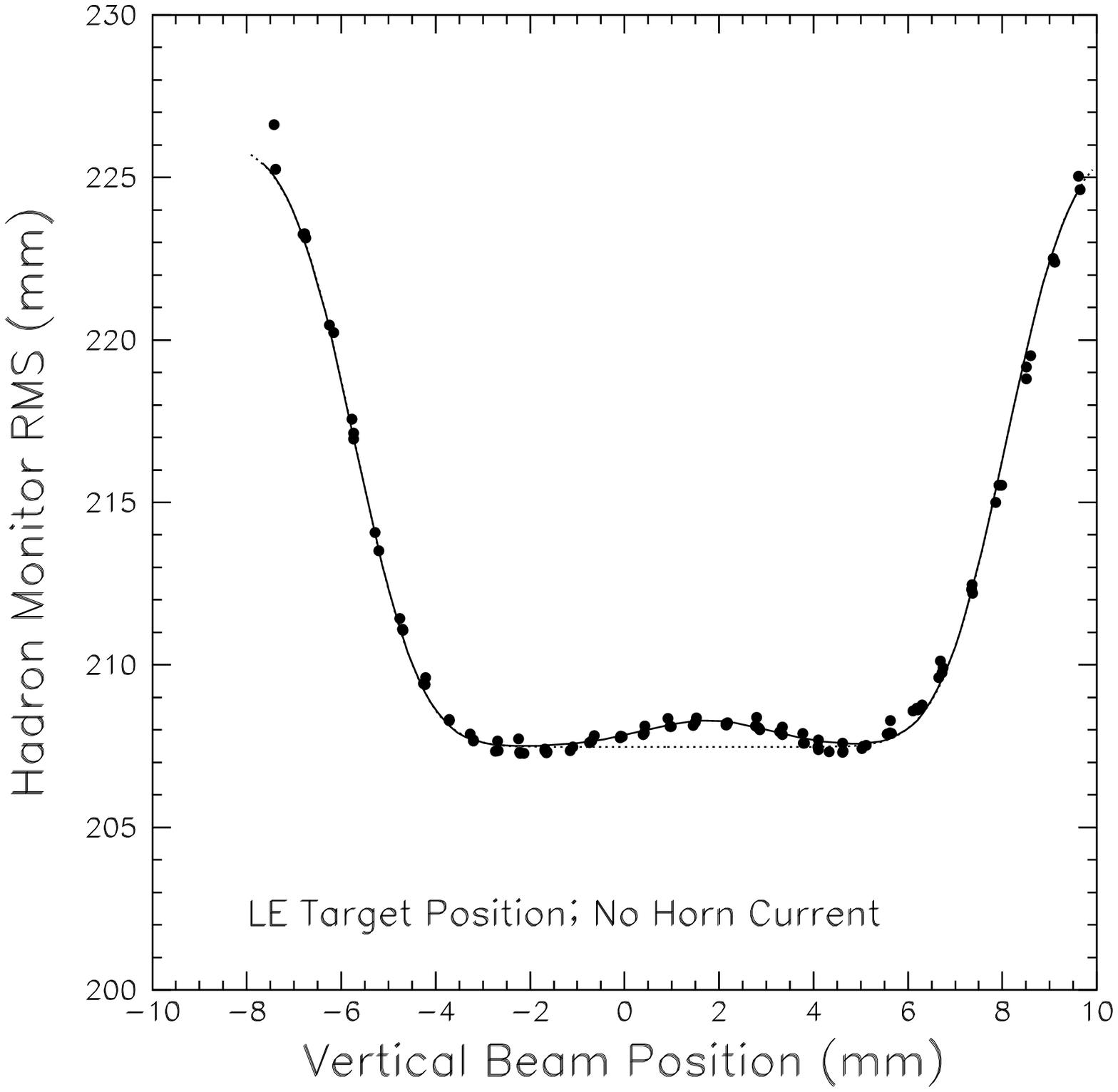}
  \vskip -.4 cm
  \caption{Vertical scans of the proton beam across the target and baffle.  
    Plotted is the vertical RMS of the 
    measured distribution in the Hadron Monitor.   The position of the 
    horizontal fin is indicated by an increase in RMS due to 
    scattering.  The solid line is a fit to a simple model
    of absorption, with the dashed line showing the fit without the 
    effect of the horizontal fin.}
  \label{fig:bmeas_vtfin_sig}
\end{figure}


A vertical scan of the proton beam across the baffle is shown in Figure~\ref{fig:bmeas_vtscans}.  The charge collected in the Hadron Monitor is reduced as the proton beam passes into the baffle due to absorption and scattering in the baffle material.  The data are fit to a constant multiplied by two error functions.  The scan indicates that the vertical position of the baffle is $(+1.2\pm0.1)$~mm relative to the center of the proton beam axis.\footnote{The later scan from April 25, 2005, after the target was re-installed in the beam line, indicated the baffle was centered at $(+0.2\pm0.1)$~mm.}

The vertical position of the target itself can be established by inferring the presence of the horizontal fin of the target.  The RMS size of the proton beam at the location of the Hadron Monitor is shown in Figure~\ref{fig:bmeas_vtfin_sig}.  The data are fit to a flat background plus two error functions for the baffle and a gaussian for the fin (the latter motivated by the proton beam size which is comparable to the size of the fin).\footnote{The increase in RMS from the baffle is somewhat saturated due to the finite angular size of the Hadron Monitor.}  The value for the centroid of the gaussian is $(+1.8\pm0.1)$~mm, approximately 1.3~mm below the nominal value of 2.26~mm (see Section~\ref{targ-fiducials})

\subsection{Horizontal Target-Baffle Measurements}
\label{bmeas_bba_tp_h}

\begin{figure}[t]
  \centering
  \vskip -.8 cm
  \includegraphics[width=13.5cm]{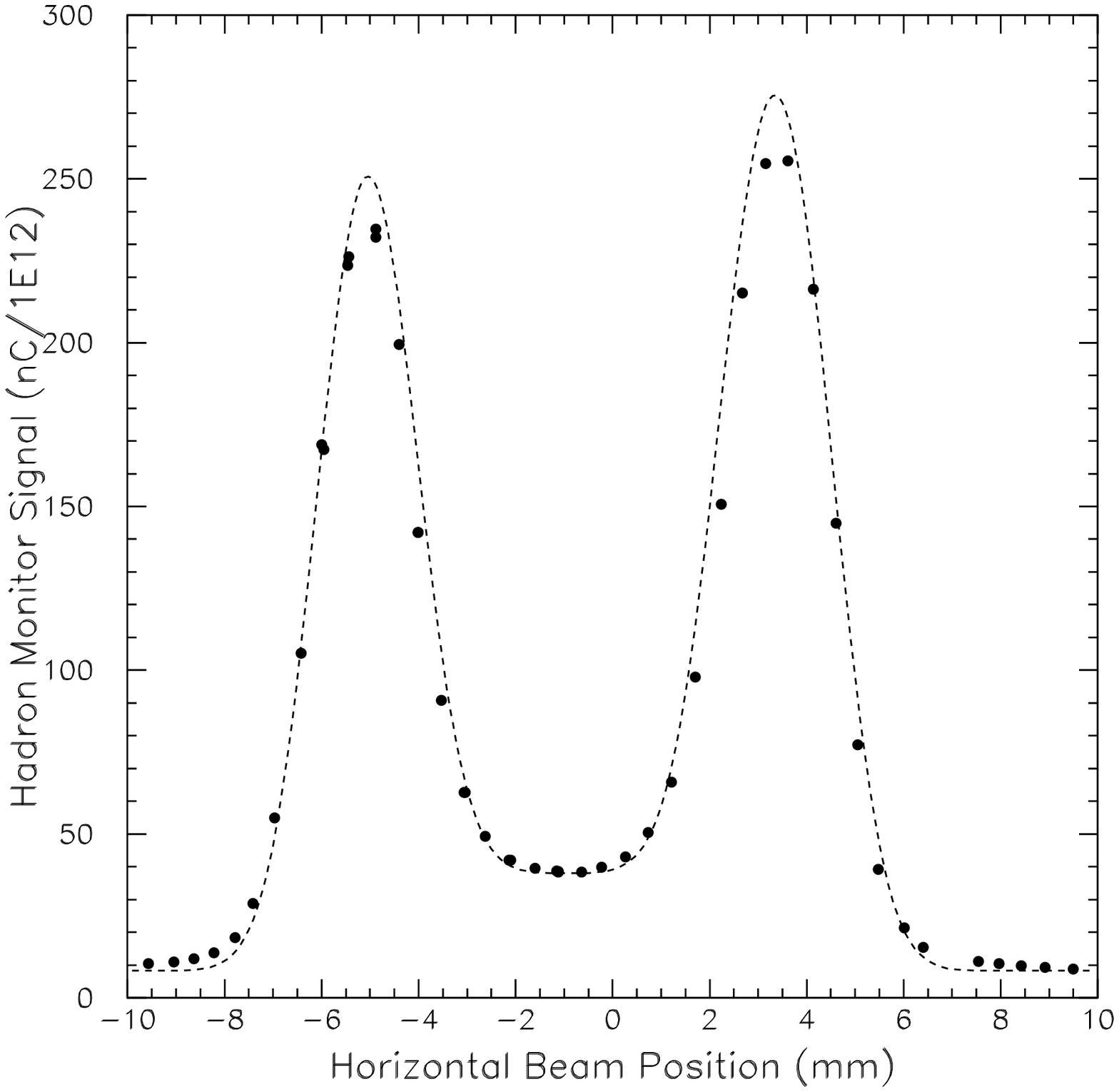}
  \vskip -.4 cm
  \caption{Data from a single horizontal scan that establishes the 
    horizontal positions of the target and baffle.
    Plotted is the total Hadron Monitor charge normalized by proton 
    intensity.  The edges of the target are found by fitting to
    the central dip; the inner edges of the baffle are found by fitting
    to the outer dips.}
  \label{fig:bmeas_htscan}
\end{figure}

A horizontal scan, shown in Figure~\ref{fig:bmeas_htscan}, is used to find the horizontal positions of the target and the baffle.  During such a scan, the beam intensity recorded by the Hadron Monitor varies due to the different amounts of material traversed:  the target is 0.96~m in length, while the baffle is 1.50~m.  When steered into the aperture between the target and baffle, much of the proton beam arrives at the Hadron Monitor unattenuated.  The data is fit to a constant multiplied by two sums of two error functions.  The target and baffle centers are fitted independently.  The fitted baffle center is at $-0.75$~mm, the target center at $-0.95$~mm.  The 0.2~mm offset between target and baffle results in the unequal signal amplitude of the two peaks corresponding to the gaps.  

The fit to the baffle width gives 10.7~mm, suggesting that it was off-center relative to the beam by as much as 1.3~mm vertically, or at an angle up to 200~$\mu$rad.  The proton beam position during this scan was $-0.5$~mm vertically, which with a 0.2~mm baffle offset, suggests the beam was off-center by only 0.7~mm, limiting the angle to 140~$\mu$rad.


\subsection{Other Indications}
\label{bmeas_bba_oth}

\begin{figure}[t]
  \centering
  \vskip -.8 cm
  \includegraphics[width=13.5cm]{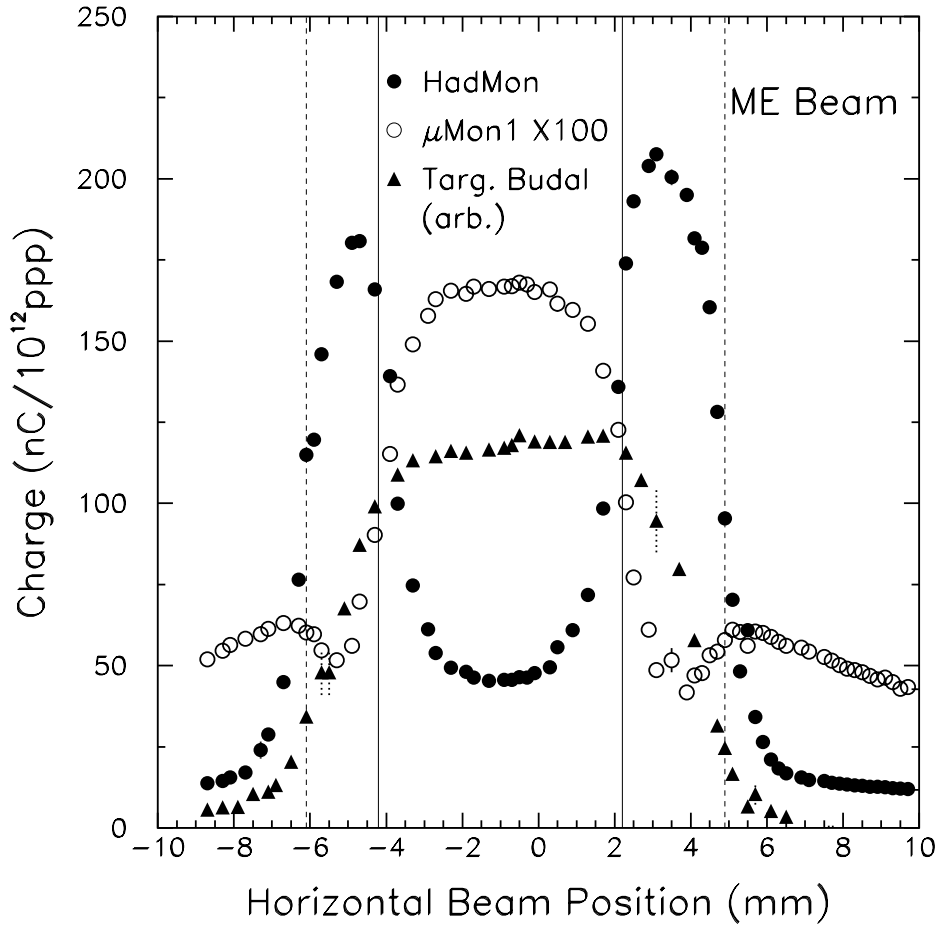}
  \vskip -.4 cm
  \caption{Data from a single horizontal scan of the target showing the pulse height seen in the Hadron Monitor, first Muon Alcove, and the target Budal signal.  The solid (dashed) lines indicate the edges of the target (baffle) as indicated by the previous beam alignment scans using the Hadron Monitor.}
  \label{fig:budalsignal}
\end{figure}

In addition to the Hadron Monitor measurements, the muon monitors and the target Budal monitors can determine the target and baffle positions.  The muon fluxes depend in a nontrivial way on how much material the proton beam traverses (see Figure~\ref{fig:le-muon-scan}), so were not used for the beam-based alignment.  The target Budal monitors were not used in the beam-based alignment because the signal was dominated by spray off the upstream baffle.  This is indicated in Figure~\ref{fig:budalsignal} by the fact that the edges of the target are poorly-resolved, and by the fact that the Budal signal is larger when the proton beam strikes just to the right of the target than to the left, consistent with the 0.2~mm misalignment between the target and baffle.  In the case of the vertical scans, furthermore, the Budal signal from the horizontal fin was very small, likely due in part to the fin's short length. 

\section{Alignment of Horns}
\label{horns}

The horizontal and vertical angles and positions of the horns were determined with a set of scans performed with the target removed and the horn currents set to zero.  All of the scans for horn alignment were performed prior to target re-installation and have the anisotropic beamspot of $\sigma_x \times \sigma_y=0.7\times1.4$~mm$^2$, resulting in somewhat better measurements in the horizontal direction. 

In these analyses the proton beam was always projected to the position of the feature in question.  However, as a single loss monitor detector is able to see multiple features, we have extrapolated the beam positions to locations that allow the comparison of these features on the same plot.  In the plots shown here the beam is extrapolated to: the Horn~1 downstream crosshair for the Horn~1 loss monitor; the Horn~2 upstream crosshair for the Horn~2 loss monitor; and the Horn~2 downstream crosshair for the Hadron Monitor RMS.  The features on the plots visible in different devices will not always line up due to the different points of extrapolation.

\subsection{Horizontal Horn Measurements}
\label{bmeas_bba_hp_h}

\begin{figure}[t]
  \centering
  \vskip -.8cm
  \includegraphics[width=13.5cm]{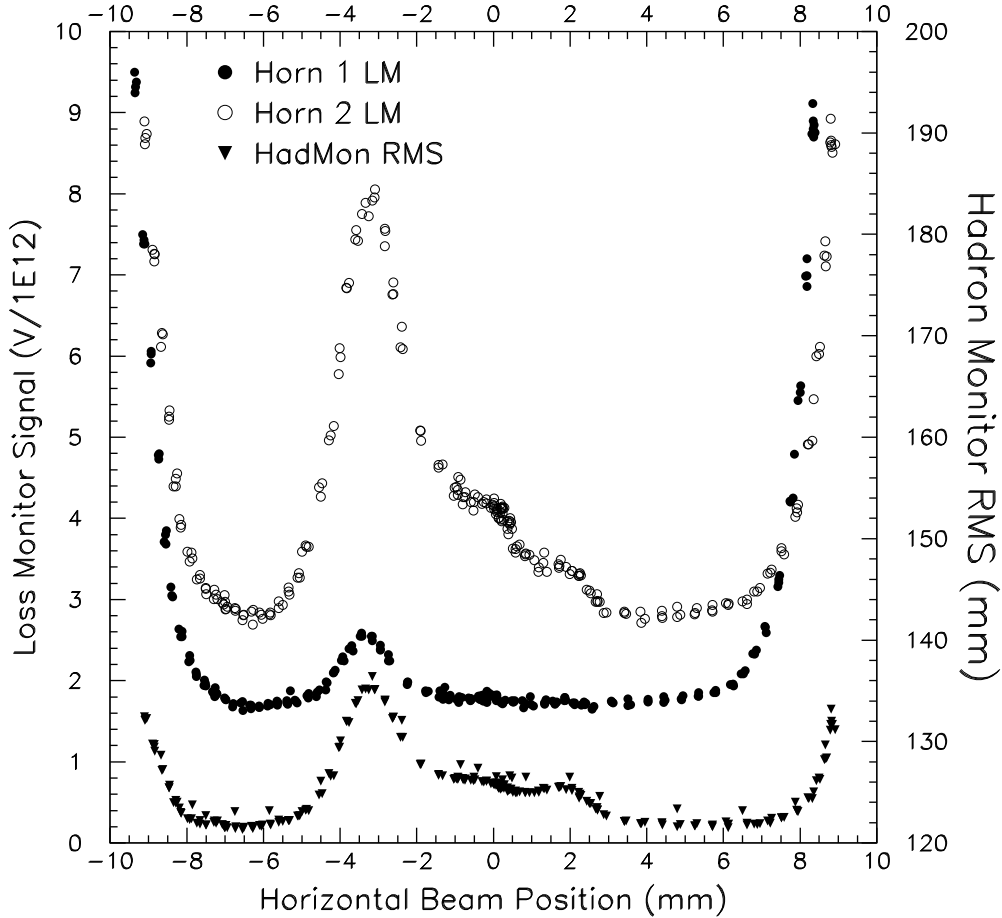}
  \vskip -.4cm
  \caption{Summary of measurements made to establish the horizontal
    positions and angles of the horns.  Shown are the signals in the
    two loss monitors, one downstream of each horn, and the vertical
    RMS of the distribution in the Hadron Monitor about the centroid
    during a horizontal scan of the proton beam across the system. }
  \label{fig:bmeas_hhscan_ped}
\end{figure}

A typical horizontal scan of the proton beam across the horns is shown in Figure~\ref{fig:bmeas_hhscan_ped}.  The data from the three detectors are plotted: ``Horn~1 LM'' is the loss monitor just downstream of Horn~1 and is sensitive to particle spray from the Horn~1 neck and the downstream crosshair on Horn~1.  ``Horn~2 LM'' is the loss monitor just downstream of Horn~2 and is sensitive to the Horn~1 neck and both crosshairs on Horn~2.  The signal from the upstream Horn~2 crosshair is about eight times as strong as that from the downstream crosshair, leading to difficulty in differentiating the signals\footnote{The Horn~1 and downstream Horn~2 crosshairs are 6~mm in length, while the upstream Horn~2 crosshair is 18~mm in length.}.  The Hadron Monitor RMS is sensitive to all material, so is not useful where crosshairs do overlap, as in Figure~\ref{fig:bmeas_hhscan_ped}.

\begin{figure}[t]
  \centering
  \vskip -.8cm
  \includegraphics[width=13.5cm]{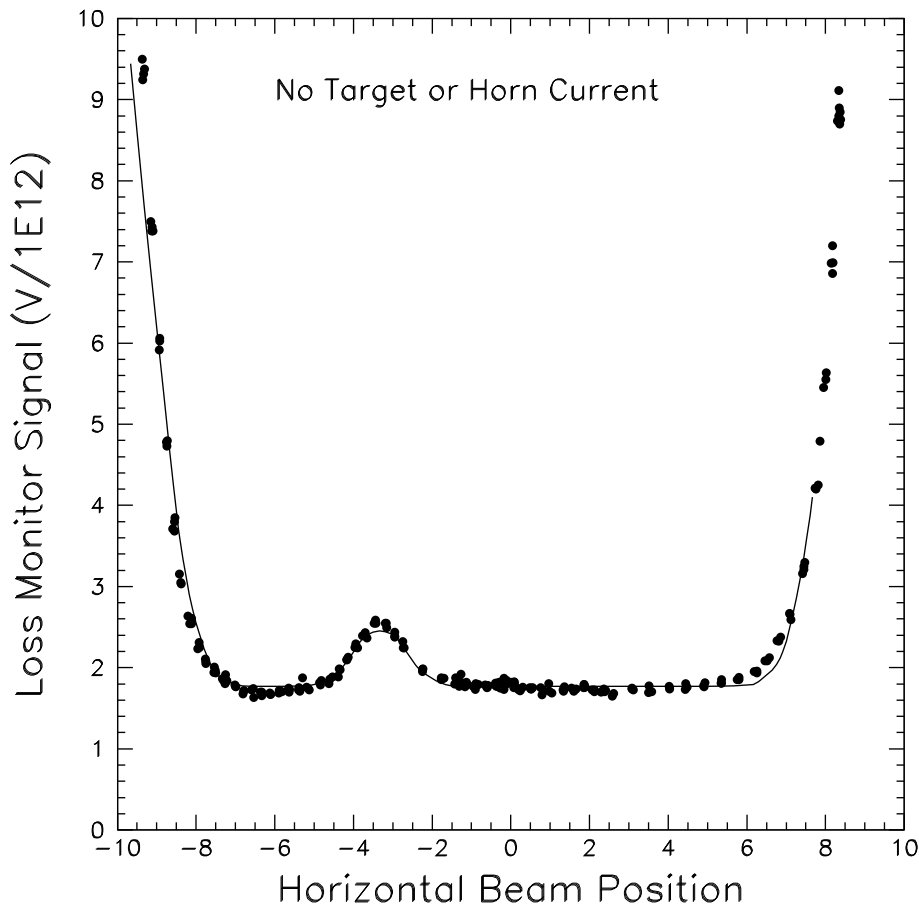}
  \vskip -.4cm
  \caption{The data from Figure~\ref{fig:bmeas_hhscan_ped} for the Horn~1
    loss monitor alone.  Superimposed on the data is the fit used to
    determine the center of the horn neck from the outer peaks,
    and the crosshair position from the central peak.}
  \label{fig:bmeas_hhscan_h1}
\end{figure}

The Horn~1 loss monitor data are shown alone on Figure~\ref{fig:bmeas_hhscan_h1} along with the fit used to determine the
horizontal position of the horn neck and crosshair.  The fit is a constant background plus two error functions for the horn neck, and a gaussian for the crosshair.  The fit results in a position of $-0.46$~mm for the horn neck and $-3.36$~mm for the downstream crosshair (which should be at 0~mm and $-2.5$~mm, respectively, if the horn was perfectly aligned to the proton beam axis).

\begin{figure}[t]
  \centering
  \vskip -.8cm
  \includegraphics[width=13.5cm]{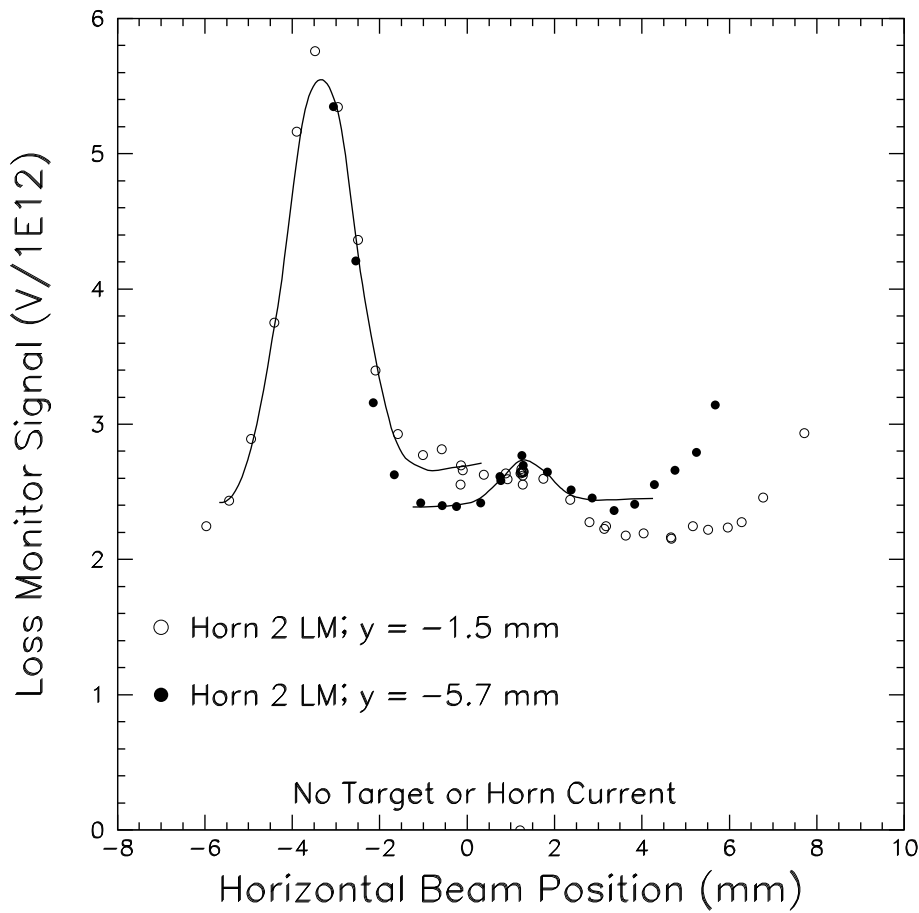}
  \vskip -.4cm
  \caption{Data from two separate horizontal scans of Horn~2 that
    establish the positions of the two crosshairs.  The two scans
    were done at different vertical positions, as indicated.  In the
    higher scan the signal from the upstream nub interferes with the 
    signal from the downstream crosshair; the second scan, displaced
    downward, avoids this problem.}
  \label{fig:bmeas_hhscan_h2}
\end{figure}

The Horn~2 crosshair signals are more difficult to separate.  The signal induced by the upstream crosshair is eight times as strong, so even the small amount of beam crossing the nub affects the signal.  The problem can be seen in the three-humped structure of the Horn~2 LM signal in Figures~\ref{fig:bmeas_hhscan_ped} and \ref{fig:bmeas_hhscan_h2}.  A similar scan is shown as the open circles in Figure~\ref{fig:bmeas_hhscan_h2}, it is displaced a fraction of a millimeter downward from the scan in Figure~\ref{fig:bmeas_hhscan_ped}, moving the proton beam out of the way of one of the horizontal nubs.  This distribution is fit to a linear background (to approximate the nub) and a gaussian.  The gaussian gives the upstream crosshair position as $-3.35$~mm (should be $-2.5$~mm).

Additional horizontal scans were performed with the beam displaced vertically so as to avoid scraping the nubs.  A second horizontal scan, with the beam displaced downward to $y=-5.7$~mm, was used to separate the downstream crosshair signal.  This scan, shown as the filled circles in Figure~\ref{fig:bmeas_hhscan_h2}, is well clear of the upstream nub, allowing the downstream crosshair to be resolved and fit to a linear background and a gaussian.  The fit gives the downstream crosshair as $+1.33$~mm (should be $+2.5~$mm).\footnote{Note that, as a result of displacing the beam downward, the Horn~1 neck appears only 7~mm across, as expected.} 

\subsection{Vertical Horn Measurements}
\label{bmeas_bba_hp_v}

\begin{figure}[t]
  \centering
  \vskip -.8cm
  \includegraphics[width=13.5cm]{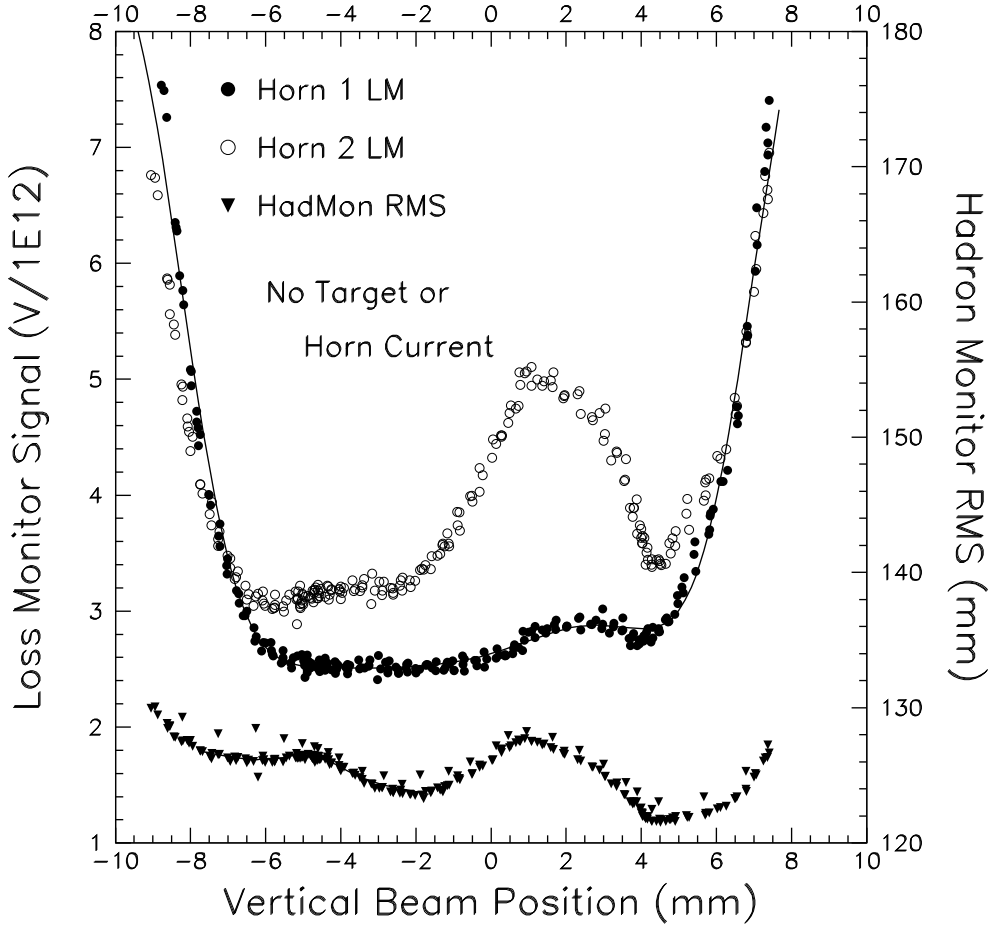}
  \vskip -.4cm
  \caption{Summary of measurements made to establish the vertical
    positions and angles of the horns.  The Horn~1 neck, crosshair nub,
    and the Horn~2 upstream nub are measured from the loss monitor signals.
    The Horn~2 downstream
    nub  position must be measured by the change in measured RMS at the
    Hadron Monitor.}
  \label{fig:bmeas_vhscan}
\end{figure}

The data from a single vertical scan of the proton beam across
the horn system are shown
in Figure~\ref{fig:bmeas_vhscan}.  Here the beam scans vertically along
the centerline of the horn neck -- avoiding the crosshair spars
as much as possible.  The signal peaks represent the horizontal nubs of the
crosshairs; the peaks are wider and less prominent than in the
horizontal scans because of the smaller amount of material in the 
nub and because of the wider vertical beam size.

The Horn~1 downstream crosshair nub and neck positions are found by 
a fit to the Horn~1 loss monitor data.  The neck, as before, is 
clearly visibly, but the nub provides only a weak bump.  Regardless,
the data is well fit by two error functions for the neck and a 
gaussian for the nub.  The fit value for the nub position is
+2.78~mm; other scans give positions of $+2.02$ and $+2.04$~mm, so an
average is used.  The fit value for the neck position is $-0.64$~mm.

The Horn~2 upstream crosshair nub is found by a fit to the peak in
the Horn~2 loss monitor data.
Two scans were fit giving positions of $+1.48$ and $+1.57$~mm.

The Horn 2 downstream crosshair nub is not resolvable in the 
Horn~2 loss monitor data. The expected position is around
$-3$~mm, and a slight bump is perhaps visible there in the loss
monitor data, but that data is dominated by background.  However,
the Hadron Monitor data can be used here as there are no other
features competing with the nub.  A fit is performed to the 
Hadron Monitor RMS with a linear background and a gaussian over 
a 6 mm range about the peak in the RMS.  The fitted value for
the position of the nub is $-4.74$~mm.

\section{Discussion of Results}
\label{summary}

The position measurements in the previous sections were used
to find the positions and some angles of the target hall
components.  
Based on fitting uncertainties and on the level of agreement between
multiple measurement samples, we estimate a position uncertainty $\pm$ 0.3 mm for the target,
baffle and Horn 1; and $\pm $0.5 mm for Horn 2.
Table~\ref{tab:bmeas_results} summarizes the transverse offsets and 
the angles of the components with respect to the beamline.  We estimate 
uncertainties of $\pm$0.2~mrad for the angles of the horns.

\begin{table}[tbp]
  \centering
  \begin{tabular}{|c|c|c|c|c|}
    \hline
	{\bf Device} & {\bf Direction} & {\bf Coord}
	& {\bf Offset (mm)} & {\bf Angle (mrad)} \\
	\hline
	Baffle & Horz & SEM & -0.75 & $<0.14$\\
	\cline{3-5}
	Baffle & Horz & BPM & -1.21 & $<0.14$\\
	\cline{2-5}
	Baffle & Vert & SEM & +0.14 &  \\
	\cline{3-4}
	Baffle & Vert & BPM & +1.12 & \\
	\cline{1-5}
	Target & Horz & SEM & -0.95 & $<0.14$\\
	\cline{3-5}
	Target & Horz & BPM & -1.41 & $<0.14$\\
	\cline{2-5}
	Target & Vert & SEM & -0.90 & \\
	\cline{3-4}
	Target & Vert & BPM & +0.13 & \\
	\hline
	Horn 1 & Horz & SEM & -0.65 & -0.18 \\
	\cline{3-5}
	Horn 1 & Horz & BPM &-1.24 & -0.18 \\
	\cline{2-5}
	Horn 1 & Vert & SEM & -0.33 & +0.20  \\
	\cline{3-5}
	Horn 1 & Vert & BPM & +0.81 & +0.26 \\
	\hline
	Horn 2 & Horz & SEM & -1.01 & -0.11 \\
	\cline{3-5}
	Horn 2 & Horz & BPM & -1.82 & -0.18 \\
	\cline{2-5}
	Horn 2 & Vert & SEM & -1.61 & -0.42 \\
	\cline{3-5}
	Horn 2 & Vert & BPM & +0.08 & -0.43 \\
	\hline
  \end{tabular}
  \caption{Positions and angles of target hall components.  
    The positions in the SEM coordinate system are calculated directly
    from the fitted values.  The beam position monitor (BPM)
    coordinate system values are derived from the SEM values, including the subsequently-measured
    differences between SEM and BPM measurements.  There is only one position
    measurement in each direction for the target and baffle, so
    there is no angle information.}
  \label{tab:bmeas_results}
\end{table}

\begin{table}[tbp]
  \centering
  \begin{tabular}{|c|c|c|c|c|c|}
    \hline
	{\bf Device} & {\bf Dir.} & {\bf Offset} & 
	{\bf Effect} & {\bf Angle} & {\bf Effect} \\
	\hline
	Baffle & Horz & -1.2 mm & 2.5\% & -0.1 mrad & $<$ 0.1\%\\
	\cline{2-6}
	Baffle & Vert & +1.1 & 2.2 & -0.7 & $<$ 0.1 \\
	\hline
	Target & Horz & -1.4 & 2.5 & -0.1 & $<$ 0.1 \\
	\cline{2-6}
	Target & Vert & +0.1 & $<$ 0.1 & -0.7 & 0.3 \\
	\hline
	Horn 1 & Horz & -1.2 & 1.1 & -0.2 & 0.3\\
	\cline{2-6}
	Horn 1 & Vert & +0.8 & 1.4 & +0.3 & 0.4\\
	\hline
	Horn 2 & Horz & -1.8 & 1.2 & -0.2 & $<$ 0.1 \\
	\cline{2-6}
	Horn 2 & Vert & +0.1 & $<$ 0.1 & -0.4 & $<$ 0.1 \\
	\hline
  \end{tabular}
  \caption{Tabulation of effects on the Far-to-Near ratio
    due to misalignments if the beam were steered at $(x,y)$ =
    (0,0)~mm.  The effects shown are the maximal distortion of the ratio of near and far detector energy spectra in the LE beam in any
    1~GeV energy bin from ($0-30$)~GeV.  The offsets are those from Table~\ref{tab:bmeas_results}
    for BPMs.  Several of the offsets, if not corrected for, produce a neutrino 
    energy spectrum distortion in excess of the 2\% limit required.  
    See Table~\ref{tab:bmeas_fneff2} for the aligned version.}
  \label{tab:bmeas_fneff1}
\end{table}

The beam-based alignment measurements were performed using the SEM's to
extrapolate the primary beam position.  During high-intensity operation of
the NuMI beam, only the BPMs are used to steer the proton beam, as the upstream 
SEM's are retracted from the beam to reduce beam loss.  Thus, it was necessary
to develop a relation between the extrapolated position determined by the 
SEMs and that determined by the BPMs.  Such was obtained and the upstream SEMs subsequently retracted.  Discrepencies
are evident in the extrapolated positions, as large as 1.7 mm,
in the vertical direction at Horn 2.  The magnitude of such discrepencies is not important given the fact that the beam axis has been 
determined from the beam-based alignment procedure.  The BPMs are
necessary only to maintain the proton beam on target.

To quantify the effect these misalignments have on the 
neutrino energy spectrum, we use a Monte Carlo simulation to find the 
beam changes due to any offset or angle.  Table~\ref{tab:bmeas_fneff1} 
summarizes the effects on the ratio of fluxes measured 
at the near and far detectors (far-to-near ratio).  While angles were not measured for the
baffle and target, we estimate an upper bound by considering
the difference in baffle and target offsets to be due to a 
common angle in their mounting.  Transverse misalignment of the target and baffle with respect
to the proton beam by 1.2~mm, for example, can cause spectral distortions exceeding 2\%.

We can correct for some of these offsets by simply redirecting the proton beam to strike
the center of the target.  
During normal beam operations, we have steered the proton beam 
at the $x = -1.2$~mm and $y = +1.0$~mm position.  In the horizontal direction, the beam is directed at our derived target center.  In the vertical direction, where centering on the target is less critical, the beam is centered on the baffle aperture so as to avoid beam halo scraping the baffle;  such scraping has been shown to distort the beam spectrum because the baffle acts as a target at an upstream location, producing higher energy neutrinos.\cite{numitdh,zwaska}  We
recalculated the target and horn offsets from the new beam location and show those in 
Table~\ref{tab:bmeas_fneff2} along with the recalculated effects
of the far-to-near ratio.  In this case, all effects contribute less
than 0.5\% distortion to the spectrum.

We have implemented a system to measure the alignment of neutrino beam elements using
the beam itself.  Such has been used to limit the systematic uncertainties for the
energy spectrum from the NuMI beam.  For the present, beam based alignment has been used
only for choosing the position at which to target the proton beam.  The 
target hall components may be eventually repositioned with respect to each 
other and angled to reduce possible effects on the neutrino flux.  

\begin{table}[tbp]
  \centering
  \begin{tabular}{|c|c|c|c|c|c|}
    \hline
	{\bf Device} & {\bf Dir.} & {\bf Offset} & 
	{\bf Effect} & {\bf Angle} & {\bf Effect} \\
	\hline
	Baffle & Horz & 0.0 mm & $<$ 0.1\% & -0.1 mrad & $<$ 0.1\%\\
	\cline{2-6}
	Baffle & Vert & +0.1 & $<$ 0.1 & -0.7 & $<$ 0.1 \\
	\hline
	Target & Horz & -0.2 & 0.4 & -0.1 & $<$ 0.1 \\
	\cline{2-6}
	Target & Vert & -0.9 & $<$ 0.1 & -0.7 & 0.3 \\
	\hline
	Horn 1 & Horz & -0.0 & $<$ 0.1 & -0.2 & 0.3\\
	\cline{2-6}
	Horn 1 & Vert & -0.2 & $<$ 0.1 & +0.3 & 0.4\\
	\hline
	Horn 2 & Horz & -0.6 & 0.2 & -0.2 & $<$ 0.1 \\
	\cline{2-6}
	Horn 2 & Vert & -0.9 & 0.4 & -0.4 & $<$ 0.1 \\
	\hline
  \end{tabular}
  \caption{Tabulation of effects on the Far-to-Near ratio
    due to misalignments if the beam were steered at $(x,y) =
    (-1.2,+1.0)$~mm -- as is the case during high-intensity beam operations.
    The effects shown are the maximal distortion of the ratio of near and far detector energy spectra in any
    1~GeV energy bin from ($0-30$)~GeV.  After steering the proton beam to the target center, none of the uncertainties
    due to offsets or angles approach the alignment budget of 2\%
    effect on the ratio of far and near detectors' energy spectra.}
  \label{tab:bmeas_fneff2}
\end{table}

\section{Acknowledgements}
It is a pleasure to thank the members of the Fermilab Accelerator Division for their support in the rapid commissioning of the NuMI beam facility.  Particularly, we would like to thank D.~Heikkinen, A.~Ibrahim, P.~Prieto, G.~Tassotto, and R.~Webber of the Accelerator Division Instrumentation Department, R. Bernstein, V. Bocean, and R. Ford of the Fermilab Alignment and Survey Group, and P.~Adamson, D.~Capista, and I.~Kourbanis of the Main Injector Department.  We thank M. Proga, R. Keisler, and M. Lang of the University of Texas for technical assistance with the primary beam SEMs and the secondary beam monitors.  We thank D. Ayres of ANL and V. Zarucheiski and V. Garkusha of IHEP-Protvino for valuable collaboration.  This work was supported by the U.S. Department of Energy under contracts DE-FG03-93ER40757 and DE-AC02-76CH3000.


\begin{thebibliography}{00}


\bibitem{numitdr} Hylen, J., {\it et al.}, ``Conceptual design for the technical components of the
        neutrino beam for the main injector (NuMI),'' Fermilab-TM-2018 (1997).
\bibitem{kopp-numi} Kopp, S., ``The NuMI Beam at Fermilab,'' in {\it Proc. 2005 IEEE Particle Accelerator Conference},
	preprint Fermilab-CONF-05-093-AD, or {\tt arXiV:physics/0508001}
\bibitem{minostdr} MINOS Collaboration, Fermilab NuMI-L-337, Oct. 1998, S. Wojcicki, spokesperson.
\bibitem{nova} NOvA Collaboration, Fermilab P929, {\tt hep-ex/0210005}, G. Feldman and M.Messier, spokespeople.
\bibitem{minerva}MINERvA Collaboration, Fermilab proposal P938, {\tt arXiv:hep-ex/0405002} (2004), J. Morfin and K. McFarland, spokespeople.
\bibitem{flexybeam} Kostin, M., Kopp, S., Messier, M., Harris, D., Hylen, J., Para, A., ``Proposal for Continuously
	Variable Neutrino Beam Energy for the NuMI Facility,'' Fermilab-TM-2353-AD (2001).
\bibitem{Casagrande} Casagrande, L. {\it et al.}, ``The alignment of the CERN West Area neutrino facility,''
	CERN Yellow Report 96-06 (1996).
\bibitem{Astier} Astier, P. {\it et al}, ``Prediction of neutrino fluxes in the NOMAD experiment'',
	{\it Nucl. Inst. Meth.}, {\bf A515}, 800-828 (2003);  {\tt arXiv:hep-ex/0306022}
\bibitem{Barish1977} Barish, S.J. {\it et al.}, ``Study of Neutrino Interactions in Hydrogen and Deuterium.
         1. Description of the experiment and Study of the Reaction
         $\nu + d \to \mu^- + p + p_s$,'' {\it Phys. Rev.} {\bf D16}, 3103 (1977).
\bibitem{numitdh} Fermilab NuMI Technical Design Handbook (2004), {\tt http://www-numi.fnal.gov/numwork/tdh/tdh\_index.html}.
\bibitem{Para2001} Para, A., ``Locating the $\nu$ Beam Elements,'' Fermilab-NuMI-B-796 (2001).
\bibitem{beamon-paper} Kopp, S. {\it et al.}, ``Secondary Beam Monitors for the NuMI Facility at FNAL,'' Fermilab-Pub-06-007-AD, submitted to {\it Nucl. Instr. Meth.} {\bf A} (2006).
\bibitem{budal} Budal, K., ``Charge transport from targets in proton beams, as a means of monitoring,'' CERN-67-17, also {\it IEEE Trans. Nucl. Sci.} {\bf 14}, 1132 (1967).
\bibitem{numitarget} Abramov, A.G. {\it et al.}, ``Beam optics and target conceptual designs for the NuMI project,'' {\it Nucl. Instr. Meth.} {\bf A485}, 209 (2002).
\bibitem{biw04}  Indurthy, D. {\it et al.}, ``Segmented Foil Secondary Emission Monitors at Fermilab,'' {\it Proc. 2005 IEEE Part. Accel. Conf.} (Fermilab-Conf-05-092-AD), also {\it AIP Conf. Proc.} {\bf 732}, 341 (2004) (Fermilab-Conf-04-520-AD).
\bibitem{sem-beamtest}  Kopp, S. {\it et al.},  ``Beam Test of a Segmented Foil SEM Grid,''  {\it Nucl. Instrum. Meth.} {\bf A554} 138 (2005).
\bibitem{prieto}  Webber, R. {\it et al.}, ``Fermilab Recycler Ring BPM Upgrade Based on Digital Receiver Technology,'' {\it AIP Conf. Proc.} {\bf 732}, 190 (2004).
\bibitem{zwaska} Zwaska, R., {\it Accelerator Systems and Instrumentation for the NuMI Beam at Fermilab},  
	PhD Thesis, University of Texas at Austin, 2005.
\end{thebibliography}
\end{document}